\documentclass[conference]{IEEEtran}
\IEEEoverridecommandlockouts

\usepackage{cite}
\usepackage{amsmath}
\usepackage{amssymb}
\usepackage{amsfonts}
\usepackage{algorithmic}
\usepackage{graphicx}
\usepackage{textcomp}
\usepackage{soul}
\usepackage{xcolor}
\usepackage{bm}
\usepackage[T1]{fontenc}

\usepackage{hyperref}
\hypersetup{
    pdftitle={Spike Talk in Power Electronic Grids -- Leveraging Post Moore's Computing Laws},    
    pdfauthor={},     
    pdfsubject={},   
    pdfcreator={},   
    pdfproducer={},  
    pdfkeywords={}, 
}

\def\BibTeX{{\rm B\kern-.05em{\sc i\kern-.025em b}\kern-.08em
    T\kern-.1667em\lower.7ex\hbox{E}\kern-.125emX}}

\begin{document}

\title{
Spike Talk in Power Electronic Grids\\ {\huge -- Leveraging Post Moore's Computing Laws}
}

\author{
\IEEEauthorblockN{}
\and
\IEEEauthorblockN{Yubo Song}
\IEEEauthorblockA{
\textit{Department of Energy} \\
\textit{Aalborg University} \\
Aalborg, Denmark \\
yuboso@energy.aau.dk}
\and
\IEEEauthorblockN{Subham Sahoo}
\IEEEauthorblockA{
\textit{Department of Energy} \\
\textit{Aalborg University} \\
Aalborg, Denmark \\
sssa@energy.aau.dk}
\and
\IEEEauthorblockN{}
}

\maketitle

\begin{abstract}
Emerging distributed generation demands highly reliable and resilient coordinating control in microgrids. To improve on these aspects, spiking neural network is leveraged, as a grid-edge intelligence tool to establish a talkative infrastructure, \texttt{Spike Talk}, expediting coordination in next-generation microgrids without the need of communication at all. This paper unravels the physics behind \texttt{Spike Talk} from the perspective of its distributed infrastructure, which aims to address the Von Neumann Bottleneck. Relying on inferring information via power flows in tie lines, \texttt{Spike Talk} allows adaptive and flexible control and coordination itself, and features in synaptic plasticity facilitating online and local training functionality. Preliminary case studies are demonstrated with results, while more extensive validations are to be included as future scopes of work.
\end{abstract}


\section{Introduction}

The energy consumption of data centers has become a major concern in modern society. As the distributed energy resources (DERs) are increasingly promoted, the carbon footprint is also becoming more critical in power grids where large amount of data are involved \cite{goldman2024ai}. Meanwhile, distributed generation is escalating the demand for coordinating control in cyber-physical microgrids to ensure operational reliability. In turn, challenges persist thereupon, involving delays \cite{sahoo2021delay} and susceptibility to cyberattacks \cite{alan2017cybermagzine}. It is therefore of much value to study on a decentralized transition of the operation paradigm to address both challenges.

Under this scenario, \textit{Talkative Power} has been accordingly developed, aiming to co-transfer power and information along transmission lines \cite{marco2023tpc}. System resilience is effectively improved, while additional energy consumption on the transmission line is inevitable. besides, as \textit{Talkative Power} relies on \textit{request-respond} information exchange protocol, its scalability is limited when multiple agents are involved in information exchange simultaneously.

In contrast, we delve into the realm of a \textit{publish-subscribe} protocol, where information is fetched locally as needed. Navigated by the biologically plausible neuron model \cite{roy2019nature, 2021loihi}, spiking neural network (SNN) has emerged with great advantage in energy efficient computation due to its event-driven feature. Beyond the von-Neumann computing architecture activated by real numbers and perceptrons, SNN leverages a leaky-charge framework instead that are triggered by asynchronous spikes.

Empowered by SNN, we formalize the \textbf{\texttt{Spike Talk}} tailored for microgrids, harnessing power flow dynamics to infer remote information locally. As spiking neurons and spiking neural networks have the features of synaptic plasticity and spike-timing-dependent plasticity (STDP), the neuromorphic infrastructure also shows potential in online learning and effectively reduces the data and energy requirements for training. Furthermore, the necessity for communication channels is eliminated, thus effectively addressing the resilience challenges in microgrids.

This article thus delineates the infrastructure of \texttt{Spike Talk} and explains it from the perspective of its decentralized and online learning features. The remaining parts of this article is organized as follows: Section II introduces the inspiration of neuromorphic infrastructure for power grids from the von Neumann bottleneck. Section III elaborates on the online learning potential of \texttt{Spike Talk} by investigating the training principles. Section IV presents a case study, and Section V concludes the entire article. By discussing its inherent advantages, more promising real-world applications are implied, which should be the future scope of our work.

\section{Post-Moore Power Electronic Grids}
\subsection{Moore's Law and Von Neumann Bottleneck}

\begin{figure}[t]
    \centering
    \includegraphics[width=0.9\linewidth]{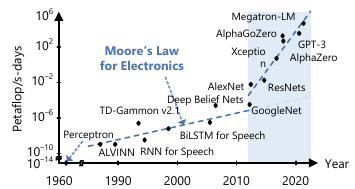}
    \caption{Development of AI models and the limit of Moore's Law. Post-Moore solutions are demanded for accommodating the new-generation AI models, with which power electronic grids involving interactive agents are facing similar challenges.}
    \label{fig_moore}
\end{figure}

The study on neuromorphic infrastructure originates from Moore's Law, which indicates that the total number of transistors on an integrated circuit (IC) should double approximately every two years. It is an empirical conclusion based on the observation of past techniques, while the implementation of computational algorithms had also been following this rule. However, the von Neumann bottleneck is observed in recent years \cite{woods2024moore}, where the room of IC development is shrinking, and the cutting-edge artificial intelligence (AI) is on the contrary calling for more powerful hardware or processing units, as shown in Fig. \ref{fig_moore} \cite{openai2018moore}. On the contrary, it is also preferred to develop more energy-efficient AI algorithms as part of the post-Moore's solutions. In power grids involving interactive agents and noticeable amount of data, Moore's Law for electronics provides similar inspirations and post-Moore solutions are also worth exploring to elevate system scalability and facilitate the processing of larger amount of data, eventually to address the concerns in energy usage and carbon emissions.

\begin{figure}[t]
    \centering
    \includegraphics[width=0.9\linewidth]{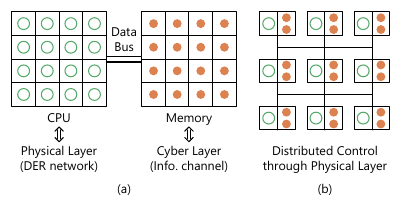}
    \vspace{-9pt}
    \caption{Illustration on the von Neumann bottleneck: (a) mapping of the von Neumann architecture to cyber-physical power grids, and (b) distributed architecture where only local data are involved.}
    \label{fig_vonneum}
\end{figure}

The basic idea is illustrated in Fig. \ref{fig_vonneum} \cite{elishai2022neuroeng, catherne2022vonneumann}. In von Neumann architecture shown in Fig. \ref{fig_vonneum}(a), the data are stored and processed in separate units, namely the memory and CPU, and data flow through the data bus between CPU and memory. In this case, the time efficiency of computational tasks are constrained by the bandwidth of the data bus, and the issues escalate when dealing with large amount of data.

The neuromorphic architecture in Fig. \ref{fig_vonneum}(b) has been consequently emerging, where data are stored close to the corresponding CPU cores. The data flows do not need to be handled by the same bus, but are distributed all over the network, thus reducing the occurrence of data congestion. This architecture can effectively mitigate the bandwidth requirement on the buses when large amount of data are demanded simultaneously, enhancing higher computational capability from hardware and thus providing possibilities to facilitate the post Moore algorithms efficiently.

\begin{figure}[t]
    \centering
    \includegraphics[width=\linewidth]{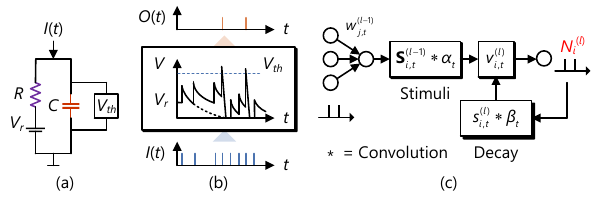}
    \caption{Illustration on the spiking neuron model: (a) equivalent \textit{RC} circuit, and (b) physics of spike generation with threshold $V_\mathrm{th}$.}
    \label{fig_spkneuron}
\end{figure}

The case in Fig. \ref{fig_vonneum}(a) is similar in conventional cyber-physical infrastructure of power grids. As per the secondary control targets, the physical layer requires remote data from the cyber layer through information channel, which limits the functionality of the system and induces the vulnerabilities of cyber attacks. Inspired by the neuromorphic architecture, a new-generation infrastructure of power grids can be formalized, the principle of which is the distributed control. Also in Fig. \ref{fig_vonneum}(b), leveraging the physical interconnections among nodes, the data bus is interpreted as the physical layer, thus requiring the local detection of remote state variables. We thereby denote this paradigm as the \textit{semantic communication} for power grids.

\subsection{Spiking Neurons and SNN-Based Power Grids}

In practice, the spiking neuron can be leveraged to implement the said semantic communication, which stems from the bio-neurons \cite{yu2017neuromorphic} and its firing paradigm dependent on the stimuli or \textit{events} accumulated on the membrane, i.e., the \textit{event-driven} paradigm \cite{sssa2024nsc, sssa2024nscnoise}. In computational neuroscience, this can be modeled into an \textit{RC} dynamic in time domain:
\begin{equation}
    I(t) = \frac{V_\mathrm{mem}(t) - V_r}{R} + C\frac{\mathrm{d}V_\mathrm{mem}}{\mathrm{d}t}
\end{equation}
as shown in Fig. \ref{fig_spkneuron}(a) \cite{yu2017neuromorphic}. Here, $I(t)$ is the input current injected to the neuron, and $V_\mathrm{mem}$ is the output membrane potential.

The spikes are generated when the threshold when the threshold $V_\mathrm{th}$ is hit, and the membrane potential is reset, e.g., through a drop by the threshold $V_\mathrm{th}$, as Fig. \ref{fig_spkneuron}(b).
\begin{equation}
    s(t) = H(V_\mathrm{mem}(t)-V_\mathrm{th})
\end{equation}
where, $H(\cdot)$ is the Heaviside step function:
\begin{equation}
    H(x)=\left\{
    \begin{array}{rcl}
    1 & & {x > 0}\\
    0 & & {x \leq 0}
    \end{array} \right.
\end{equation}

Considering the interconnection of spiking neurons, spike response model (SRM) has also been established for spiking neural networks (SNN) \cite{roy2019nature}, as shown in Fig. \ref{fig_spkneuron}(c). Each neuron $N_{i,l}$ establishes connections with its preceding neurons via synaptic weights ${w_{j,\:l-1},\: \forall j\in \mathrm{Layer}\: (l-1)}$. When a spike is received from the preceding neurons, the membrane potential $v_{i,\:l}$ will experience a momentary increase (stimuli) and subsequently leakage of charge, namely the leaky-integrate-and-fire (LIF) model. An output spike is thereby set off given that the input membrane potential surpasses the threshold $V_\mathrm{th}$:
\begin{subequations}
\begin{align}
    v^{(l)}_{i,\:t} &= \sum_{j=1}^{N}w^{(l-1)}_{j}\cdot(\alpha_t*s^{(l-1)}_{j,\:t}) + \beta_t*s^{(l)}_{i,\:t};\:\:\:\: s^{(l)}_{i,\:t} \\
    &= H(v^{(l)}_{i,\:t}-V_\mathrm{th})
\end{align}
\end{subequations}

In conventional scenarios, SNN can be applied similar to other types of neural networks. For instance, in microgrid applications, we encode the voltage $v_k$ and current $i_k$ at each node into spikes and train the neural network accordingly. In \cite{yu2017neuromorphic}, several encoding methods are elaborated, including rate coding (higher sampled value mapped into larger number of spikes) and latency coding (higher sampled value mapped into earlier presence of spikes), etc. By defining the events via dynamic thresholds, event-driven semantic sampling is thereby implemented. 

In our previous work \cite{sssa2024neuroac}, we extract the most significant data and trigger the binary events based on physical transients. When a dynamic change occurs in, e.g., the sampled voltage $v$, its corresponding event $\Omega_{v}$ is triggered with a transition from \texttt{0} to \texttt{1}, and returns to \texttt{0} eventually when a new steady state is reached. A similar approach is applied to detect dynamics in sampled current $i$ (or other state variables), and the events are denoted as $\Omega_{i}$. The event $\Omega$ for semantic sampling and data collection is subsequently defined as:
\begin{equation}
    \label{eq_event}
    \Omega =\{\Omega_{v}  \quad  \texttt{OR} \quad  \Omega_{i}\}
\end{equation}
Using (\ref{eq_event}), $\{v_k,i_k \ \forall k \in N_\mathrm{DER}\}$ of inverter $k$ will only be sampled for inputs to the inferential learning of SNN when there is an event, or $\Omega$ is \texttt{1}.

\subsection{Synaptic Plasticity for Neuromorphic Power Grids}

\begin{figure}[t]
    \centering
    \includegraphics[width=\linewidth]{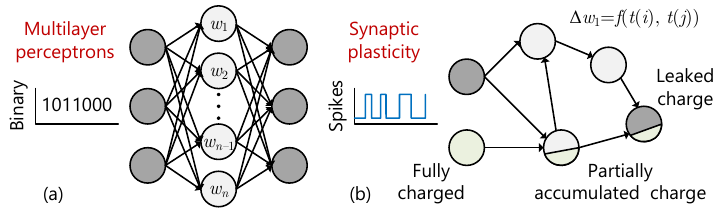}
    \vspace{-9pt}
    \caption{Going beyond von-Neumann computing architecture for exchange of information between neurons: (a) Binary-activated NN trained using multilayer perceptron policy with iterative learning based weight updates, (b) spiking neural networks using synaptic plasticity allows online learning using the spike-based events from power flows at each node.}
    \label{fig_chargemdl}
\end{figure}

Moreover, this neuron model and the stated interconnection scheme also enable the neuromorphic architecture with noticeable synaptic plasticity beyond the von Neumann computing architecture, as illustrated in Fig. \ref{fig_chargemdl}. Conventional perceptrons are interconnected in layers, the training of which requires iterations for weight updates. In comparison, the spiking neurons are connected as a network, like the physical layer in power grids. The state of a neuron can be fully / partially charged or leaked, which are dependent on the spikes received from preceding neurons, enabling online learning and showing synaptic plasticity in practice.

\begin{figure}[t]
    \centering
    \includegraphics[width=0.9\linewidth]{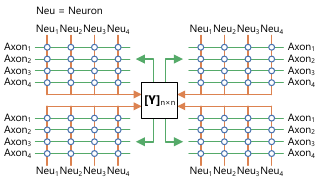}
    \vspace{-9pt}
    \caption{A system-level view of the connectivity of local neuromorphic networks. Local neural networks are coupled through the system admittance matrix $\mathbf{[Y]}_{n\times n}$.}
    \label{fig_nnsystem}
\end{figure}

This forms the theoretical basis of \textit{neuromorphic power grids}. Exemplifying a DC microgrid consisting of distributed energy resources (DERs) interfaced by power electronics converters, we thereby outline the following features to describe this concept \cite{sssa2024spktalk}:
\begin{enumerate}
    \item The entire system can be mapped into a neural network, with each converter being a single neuron. The RC dynamic in LIF model matches the impedance network dominated by resistances and capacitances.
    \item When the weights are set aligned with the system impedances / admittances, the neural network is essentially mirroring the behaviors of the system. This indicates the possibility of using only local sampling to detect the remote states for secondary controllers.
    \item With the synaptic plasticity, it is possible to update the weights locally by integrating the local sampling and control targets, which is the potential of online learning. In this way, each local neural network or each converter is "learning" from the inherent system dynamics.
\end{enumerate}

\section{Spike Talk for Power Electronic Grids}
\subsection{From Neuromorphic Infrastructure to Spike Talk}

In power electronic dominated grids, we justify the name of this paradigm as \texttt{Spike Talk}, as spike neurons are utilized to convey the system states \cite{sssa2024spktalk}. If the aforementioned principles are followed, then the spiking neural networks are coupled through the system admittance matrix $\mathbf{[Y]}_{n\times n}$ (the physical layer), as depicted in Fig. \ref{fig_nnsystem}. Though the neural networks are executed locally, the physical coupling can contribute to maintaining the consistency of weight changes during online learning as well as the convergence of gradients regarding the loss functions.

\begin{figure}[t]
    \centering
    \includegraphics[width=\linewidth]{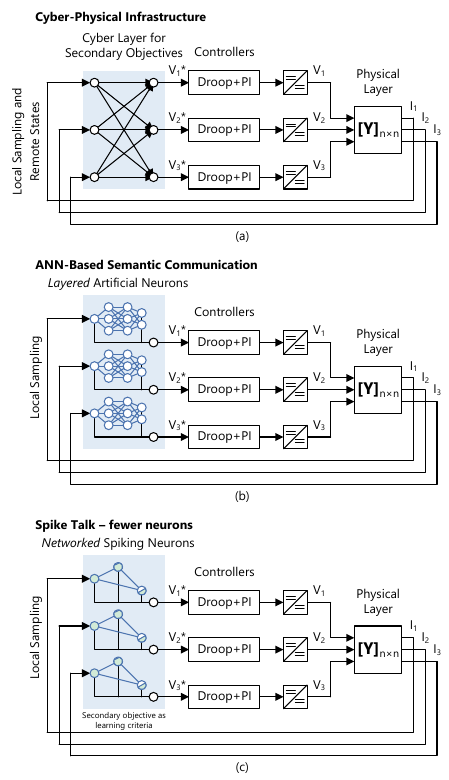}
    \vspace{-9pt}
    \caption{Comparison on the signal flows and requirements on data processing: (a) conventional cyber-physical infrastructure, (b) ANN-based semantic communication, and (c) \texttt{Spike Talk} which shows advantages in both system resiliency and energy efficiency for data processing.}
    \label{fig_semantic}
\end{figure}

We inspect a simple scenario of a DC grid consisting of interconnected DC-DC converters and governed by droop relationships. The proposed \texttt{Spike Talk} is then compared in Fig. \ref{fig_semantic}, with conventional the cyber-physical infrastructure and the semantic communication using basic artificial neural network (ANN) \cite{sssa2024nsc}:
\begin{enumerate}
    \item Cyber-physical infrastructure in Fig. \ref{fig_semantic}(a) requires both local sampling and the detection of remote states, which embeds the vulnerability to cyber attacks. As the cyber layer is essential in this case for data transfer, the system controllability is also limited by the capacity of the data buses all the time, which accords with the disadvantages of the von Neumann architecture mentioned in Fig. \ref{fig_vonneum}.
    \item In semantic communications in Fig. \ref{fig_semantic}(b), the cyber layer is dismissed, improving the system resilience against cyber attacks. However, the remote data is still required for initializing training of ANN, and the energy efficiency during both training and execution is high due to the large number of perceptrons connected in layers.
    \item \texttt{Spike Talk} in Fig. \ref{fig_semantic}(c) further goes beyond the ANN-based semantic communications. The networked structure can not only mirror the physical layer and simplify the initialization, but also enable the online training for higher adaptability to, e.g., various potential fault scenarios in power grids. Besides, as the number of neurons accords with that of the nodes, the neural networks are significantly simplified, implying enhanced energy efficiency in practice.
\end{enumerate}

\subsection{Principles of Hebbian-Disciplined Spike Talk}

Spiking neural networks can be trained by general back-propagation approaches, while we tend to emphasize its capability of Hebbian learning, or online learning as per the control targets. This feature extends the \textit{semantic communication} among converters to the \textit{comprehension} of the system dynamics, promoting the adaptability of decentralized control for power grids.

\begin{figure}[t]
    \centering
    \includegraphics[width=0.9\linewidth]{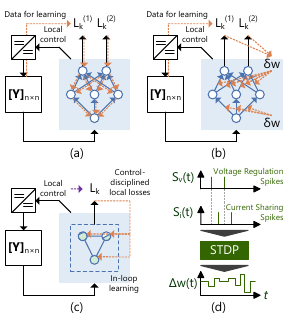}
    \vspace{-9pt}
    \caption{Learning principles for the layered or networked SNN-based infrastructures \cite{jason2023training}, which includes: (a) gradient back-propagation and (b) perturbation learning for layered infrastructures, and (c) local losses for networked infrastructures, where $L_k$ represents the local loss functions of the $k$-th node, and the superscripts in (a) and (b) represent the indices of neurons in the output layer. Online in-loop learning can be implemented by initializing the synaptic weights and leveraging the control-disciplined local losses, e.g., (d) based on STDP.}
    \label{fig_learning}
\end{figure}

Several training principles are shown in Fig. \ref{fig_learning}(a)-(c) \cite{jason2023training}. For layered structure, the gradient back-propagation in (a) is the most basic training method, where the gradient of loss functions are leveraged to update the weights. Considering the feedback path provided by the physical layer, the  perturbation learning in (b) is also an option. The gradient of back-propagation-based training can be calculated based on the following equation \cite{jason2023training}:
\begin{equation}
    \cfrac{\partial L}{\partial W_\mathrm{in}} = \cfrac{\partial L}{\partial S_\mathrm{out}} \cdot \cfrac{\partial S_\mathrm{out}}{\partial U} \cdot \cfrac{\partial U}{\partial W_\mathrm{in}}
\end{equation}
where $L$ is the data-based loss function, $W_\mathrm{in}$ is the input weight, $S_\mathrm{out}$ is the output spike, and $U$ is the generated membrane potential. The first and third term can be directly calculated, while the second term should be approximated as surrogate gradient through:
\begin{equation}
    \cfrac{\partial \Tilde{S}}{\partial U} = \cfrac{1}{\pi} \cdot \cfrac{1}{1+(\pi U)^2}
\end{equation}

However, in these two cases, secondary control is normally generated from the remote states estimated by the neural network, thus remote data are still required for training.

When the neurons are connected in networked structure, another method becomes possible utilizing local losses. Since in DC grids, droop control is essentially an additional damping resistance, the secondary control targets can be merged into the changing weights as control-disciplined local losses. For instance, the spike-timing-dependent plasticity (STDP) feature of SNN introduced in \cite{yu2017neuromorphic} can be leveraged to account for secondary control objectives. An example is given in \cite{sssa2024spktalk}, to implement both voltage regulation and power management by aligning the voltage and current spikes, as illustrated in Fig. \ref{fig_learning}(d). The modification of droop gain is federated by the secondary control objectives:
\begin{enumerate}
    \item Voltage follows the global average reference;
    \item Current follows the power sharing requirements;
    \item When voltage and current are coded through rate and latency coding, respectively, the control target is met when the cross entropy of both reaches minimum.
\end{enumerate}
Thus, the droop gain can be calculated as:
\begin{equation}
    R_\mathrm{d}' = R_\mathrm{d} - a\cdot\Delta w(t)
\end{equation}
where, $\Delta w(t)$ is the variation of weights in the SNN.

\subsection{Key Features of Spike Talk}

Therefore, \texttt{Spike Talk} also shows the following features based on the aforementioned functionalities:
\begin{enumerate}
    \item \textit{Energy efficiency}: As \texttt{Spike Talk} roots in the event-driven paradigm, the neurons only show an state change when there is an event or transient. Also, as the desired neural network is highly consistent with the physical layer, it is not necessary to sample large amount of data for training and execution. Hence, it can effectively address the issue in energy efficiency of the data centers in power electronics grids. 
    \item \textit{Publish-subscribe paradigm}: \texttt{Spike Talk} requires local controllers to learn from the inherent system dynamics, without requiring data from remote nodes, which can be recognized as the \textit{publish-subscribe} paradigm of the aforementioned neuromorphic infrastructure, and different from the conventional \textit{request-receive} paradigm relying on the cyber layer. This feature meets the requirement on distributed computation and turns out to be a favorable post-Moore solution for power grids.
\end{enumerate}

\section{Performance Demonstrations}
\begin{figure}[t]
    \centering
    \includegraphics[width=0.9\linewidth]{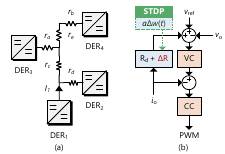}
    \caption{Configuration of the study case: (a) topology as a DC microgrid consisting of four nodes (DERs), and (b) droop-based controller at each node.}
    \label{fig_testcase}
\end{figure}

A case study has been conducted by simulation to demonstrate the features of \texttt{Spike Talk}. The studied system is shown in Fig. \ref{fig_testcase}, which is a droop-based DC microgrid consisting of four nodes. The secondary controllers aim at regulate the average voltage and the power sharing, and the key parameters are listed in appendix.

\begin{figure}[t]
    \centering
    \includegraphics[width=0.9\linewidth]{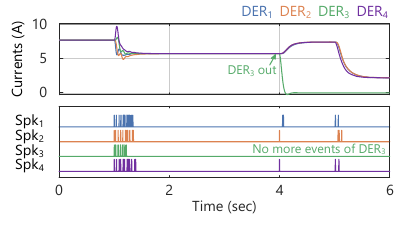}
    \vspace{-9pt}
    \caption{Simulation results showcasing the event capturing and spike generation during regular transient and DER outage.}
    \label{fig_results1}
\end{figure}

We first test the event capturing for SNN in Fig. \ref{fig_results1}. We utilize rate coding, and the events are captured only when there is a transient. When $\mathrm{DER}_3$ is out at \textit{t} = 4 s, the spikes of $\mathrm{DER}_3$ are no longer generated, or in other words, the neuron representing $\mathrm{DER}_3$ is no longer active. This justifies the basis for SNN to adapt fault scenarios in grids or time-dependent system architectures.

\begin{figure}[t]
    \centering
    \includegraphics[width=0.9\linewidth]{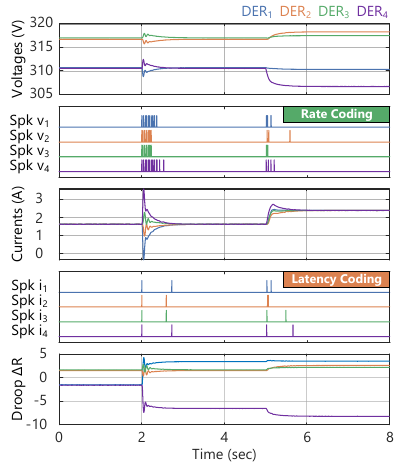}
    \vspace{-9pt}
    \caption{Simulation results showcasing the generated voltage and current spikes, and the change of droop coefficients as per the STDP-governed training objective specified in Section III-B.}
    \label{fig_results2}
\end{figure}

We further showcase the STDP feature specified in Section III-B, and the results are shown in Fig. \ref{fig_results2}. With the initial weights are assigned, the spikes for voltage and current are captured by rate coding and latency coding, respectively, and the change of droop coefficients is recorded. From the results, the spikes for voltage and current are generated based on different coding methods to implement STDP. With this, the droop gain can be adjusted adaptively, which is the main purpose of online learning. Nevertheless, the results only involve preliminary cases, and we plan to extend the demonstrations by online experiments as future scope of work.

\section{Conclusions and Future Perspectives}

This paper sheds light on \texttt{Spike Talk}, a novel concept by employing spiking neural network (SNN) for semantic communication in power electronics grids, from the perspective of its decentralized and online learning features. Leveraging STDP and the networked architecture of SNN, \texttt{Spike Talk} dismisses the reliance on physical information channels and shows the potential in enhancing system energy efficiencies. As the validation of this concept is still not comprehensive, we aim to showcase \texttt{Spike Talk} in more scenarios and performing online experimental tests, to further evaluate its potential advantages in piloting future applications.

\appendix
\textit{Key Parameters of the Study Case}: $V_\textrm{ref}$ = 315 V, $C_1$ = 450 \textmu F, $C_2$ = 500 \textmu F, $C_3$ = 480 \textmu F, $C_4$ = 520 \textmu F, $r_\textrm{a}$ = 0.5 $\Omega$, $r_\textrm{b}$ = 0.25 $\Omega$, $r_\textrm{c}$ = 0.6 $\Omega$, $r_\textrm{d}$ = 0.8 $\Omega$, $R_\textrm{d}$ = 2, a = 2.

\vfill

\bibliographystyle{IEEEtran}
\bibliography{bibliography}

\end{document}